\documentstyle[psfig,preprint,aps]{revtex}

\tightenlines

\def\beq{\begin{equation}}
\def\eeq{\end{equation}}
\def\bea{\begin{eqnarray}}
\def\eea{\end{eqnarray}}

\begin{document}
\draft

\title{Photon-Photon Interaction in a Photon Gas} 

\author{Markus H. Thoma\footnote{Heisenberg fellow}}
\address{Theory Division, CERN, CH-1211 Geneva, Switzerland\\ 
and\\
Institut f\"ur Theoretische Physik, Universit\"at Giessen, 35392 
Giessen, Germany}

\maketitle

\begin{abstract}
Using the effective Lagrangian for the low energy photon-photon
interaction the lowest order photon self energy 
at finite temperature and in non-equilibrium is calculated
within the real time formalism.
The Debye mass, the dispersion relation, the dielectric tensor,
and the velocity of light following from the photon self energy are discussed.
As an application 
we consider the interaction of photons with the cosmic microwave background
radiation.
 
\end{abstract}

\bigskip


\newpage

Photon-photon scattering has been considered already a long time ago
\cite{Eul36}. To lowest order QED perturbation theory it is caused by
the so-called box diagram, which contains an electron loop.
For low energy photons, i.e. for center of mass energies below
the threshold for $e^+$-$e^-$ pair creation, an effective Lagrangian
for the photon-photon interaction has been derived by integrating out the 
electrons\cite{Sch51}, 
\beq
{\cal L}_I= a\> (F_{\mu \nu} F^{\mu \nu})^2
+b\> F_{\mu \nu} F^{\nu \rho}
F_{\rho \sigma} F^{\sigma \mu},
\label{eq1}
\eeq
where $F_{\mu \nu}=\partial _\mu A_\nu -
\partial _\nu A_\mu$ is the electromagnetic field strength tensor and
\beq
a=-\frac{5\alpha^2}{180m_e^4},
\;\;\;\;\;\;\;\;\;\;\;\;\;\;\;\;\;\;\;\;\;\;\;\;\;
b= \frac{7\alpha^2}{90m_e^4}
\label{eq2}
\eeq
with the fine structure constant $\alpha \simeq 1/137$ 
and the electron mass $m_e$.
This effective Lagrangian containing an effective 
4-photon interaction describes the deviation from the classical Maxwell 
theory by quantum effects. The 4-photon vertex in momentum space
following from this 
Lagrangian has been derived only recently \cite{Hal93}.  The coupling
constant corresponding to this vertex is of the order $\alpha^2/m_e^4$.

The lowest order photon self energy $\Pi_{\mu \nu}$, 
which is given by the tadpole diagram 
of Fig.1, vanishes at zero temperature after dimensional regularization
\cite{Hal93}. However, at finite temperature tadpole diagrams lead to
a finite result \cite{Kap89}. The in-medium photon self energy determines
the Debye screening, the photon dispersion relations, and the dielectric
functions of the system. Some of the results presented here are already
discussed in the literature \cite{Wol83,Kon98}. 
Here we want to treat the photon self energy and its consequences
in a systematic and comprehensive way starting from the real time formalism, 
which also allows an extension to non-equilibrium situations. 
As an application
we consider the influence of the 
cosmic microwave background radiation on low energy photons.
Although medium effects of the photon gas are expected to be very small 
due to the extremely weak photon-photon coupling at low energies,
these effects might be interesting after all since the properties of
the photons are experimentally very well known. For example
there are very restrictive
upper limits for the photon mass \cite{Par98}, namely $m_\gamma < 
2 \times 10^{-16}$ eV measured in the laboratory and $m_\gamma < 
10^{-27}$ eV using arguments about the galactic magnetic field.
These upper limits follow from searching for violations of the Maxwell
theory, in particular of the Coulomb law \cite{Jac99}. Hence these 
limits should be compared to the Debye mass caused by the background
of thermal particles. It should be noted that the Debye mass, which 
follows from the photon self energy, does not violate gauge symmetry of QED.

Mass effects due to charged particles of the
thermal background, of which the electron is the lightest, are 
suppressed exponentially by a factor $\exp (-m_e/T)$, where $T=2.7$ K
is the temperature of the background \cite{Nie83}. This argument, however, 
does not hold in the case of the photon-photon interaction according to 
(\ref{eq1}), since the electrons, which have been integrated out, come 
from vacuum polarization.
Therefore it appears to be worthwhile to reconsider the effective
photon mass due to the cosmic background radiation.
Further interesting quantities following from the photon self energy are
the dispersion relations
of photons in a thermal photon gas and its dielectric function, related to 
the index of refraction and the velocity of light.

Using the notation $P=(p_0,{\bf p})$ and $p=|{\bf p}|$ the retarded
photon self energy according to Fig.1 reads
\beq
\Pi_{\mu \nu}(P)=-\frac{1}{2}\> \int \frac{d^4Q}{(2\pi )^4}\> 
D^{\rho \sigma }(Q)\>
\Gamma _{\rho \mu \sigma \nu}(Q,P,Q,P).
\label{eq3}
\eeq
where $\Gamma _{\rho \mu \sigma \nu}$ is the effective 4-photon vertex
and the factor $1/2$ a symmetry factor associated with the tadpole diagram.
Adopting the real time formalism in the Keldysh representation \cite{Car99}
the photon propagator in a general covariant gauge with gauge 
parameter $\xi $ reads
\beq
D^{\rho \sigma}(Q)= -\left (g^{\rho \sigma}-\xi \frac{Q^\rho Q^\sigma}{Q^2}
\right )\> \frac{1}{2} \left [D_R(Q)+D_A(Q)+D_F(Q)\right ].
\label{eq4}
\eeq
The retarded (R), advanced (A), and symmetric (F) propagators are given
by
\bea
D_{R,A}(Q)&=&\frac{1}{Q^2\pm i\, {\rm sgn}(q_0)\varepsilon},\nonumber \\
D_F(Q)&=&-2\pi i\> [1+2n_B(Q,x)]\> \delta(Q^2)
\label{eq5}
\eea
whith the non-equilibrium photon distribution $n_B$ depending on the 
momentum and the space-time coordinate. In equilibrium it reads
$n_B^{\rm eq}=1/[\exp(|q_0|/T)-1]$. In the following we restrict ourselves
to isotropic momentum distributions\footnote{The 
anisotropic case has been discussed in Ref.\cite{Mro00} for QED and QCD.}, 
i.e. $n_B=n_B(q_0,q,x)$.
Then there are only two independent components of $\Pi_{\mu \nu}$, which 
depend on $p_0$ and $p$ \cite{Kap89}. For these components we choose
\bea
\Pi_L(p_0,p)&=&\Pi_{00}(P),\nonumber \\
\Pi_T(p_0,p)&=&\frac{1}{2}\> \left (\delta_{ij}-\frac{p_ip_j}{p^2}
\right )\> \Pi_{ij}(P).
\label{eq6}
\eea
It should be noted that the longitudinal component is sometimes defined
differently \cite{Kap89}.

Since the zero temperature contributions vanish we find
\bea
\Pi_L(p_0,p)&=&-\frac{i}{2}\> \int \frac{d^4Q}{(2\pi )^3}\> 
n_B(q_0,q,x)\> \delta (Q^2)\> {\Gamma ^{\rho}}_{0 \rho 0}(Q,P,Q,P),\nonumber \\
\Pi_T(p_0,p)&=&-\frac{i}{4}\> \left (\delta_{ij}-\frac{p_ip_j}{p^2}\right )
\int \frac{d^4Q}{(2\pi )^3}\> 
n_B(q_0,q,x)\> \delta (Q^2)\> {\Gamma ^{\rho}}_{i \rho j}(Q,P,Q,P).
\label{eq7}
\eea
Adopting the expression for the 4-photon vertex given in Ref.\cite{Hal93},
where some factors of 2 \cite{Dic98} and the sign are corrected, we obtain
\bea
&& {\Gamma ^{\rho}}_{0 \rho 0}(Q,P,Q,P)\Biggl |_{Q^2=0}
=16i\> (4a+3b)\> [({\bf p}\cdot {\bf q})^2-p^2q^2],\nonumber \\
&& \left (\delta_{ij}-\frac{p_ip_j}{p^2}\right ) {\Gamma ^{\rho}}_{i \rho j}
(Q,P,Q,P)\Biggl |_{Q^2=0}=-16i\> (4a+3b)\> (p_0^2+p^2)\> q^2\> 
\left[1+\frac{({\bf p}\cdot {\bf q})^2}{p^2q^2}\right ].
\label{eq8}
\eea
In (\ref{eq7}) the gauge fixing parameter $\xi $ does not appear since
$Q^\rho Q^\sigma \Gamma_{\rho \mu \sigma \nu}(Q,P,Q,P)=0$ as can be
shown explicitly. Hence the lowest order photon self energy is gauge
invariant.

Inserting (\ref{eq8}) into (\ref{eq7}) we end up with the final result
\bea
\Pi_L(p_0,p)&=&-\gamma\> p^2,\nonumber \\
\Pi_T(p_0,p)&=&-\gamma\> (p_0^2+p^2),
\label{eq9}
\eea
where
\beq
\gamma=\frac{44}{135\pi^2}\> \frac{\alpha^2}{m_e^4}\> \int_{0}^{\infty} dq\>
q^3\> n_B(q,x).
\label{eq10}
\eeq
In equilibrium (\ref{eq10}) reduces to
$\gamma = (44\pi^2/2025)\> \alpha^2\> (T/m_e)^4$. Then (\ref{eq9}) agrees 
apart from the sign for $\Pi_T$ with Ref.\cite{Kon98}.

Now we want to discuss the physical consequences of our result. First we
consider Debye screening in the photon gas. It has been argued 
that there is no Debye mass due to the photon-photon interaction 
since the corresponding vertex vanishes if one of the external
legs has zero momentum \cite{Wol83}. However, this argument is based on an
inconsistent definition for the Debye mass 
\beq
m_D^2=\Pi_{L}(p_0=0,p\rightarrow 0),
\label{eq11}
\eeq
where $p=|{\bf p}|$.
This definition can lead to gauge dependent results and is not  
renormalization-group invariant \cite{Reb93}. Instead of (\ref{eq11})
the Debye mass should be determined self consistently from the pole
of the longitudinal photon propagator, i.e. from \cite{Reb93}
\beq
m_D^2-\Pi_{L}(p_0=0,p^2=-m_D^2)=0.
\label{eq12}
\eeq
Only if $\Pi_{L}(p_0=0,p)$ does not depend on $p$, the definition 
(\ref{eq11}) agrees with (\ref{eq12}). Although this is not the case
here, the Debye mass following from (\ref{eq9}) and (\ref{eq12})
vanishes also. This means that the photon-photon interaction in a
photon gas does not lead to a screening of the Coulomb potential.
Of course, there is also no static magnetic screening.

As another point we mention that the tadpole self energy of Fig.1 has no
imaginary part, i.e. neither real nor virtual photons are damped
to this order. In QED damping arises from the box diagram only above
the threshold for electron-positron pair production. This effect, however,
is not included in the effective theory (\ref{eq1}) for low energy
photons. Damping will be present in the effective theory
at the two-loop level (sunset diagram) corresponding to 
photon-photon scattering.

Next we discuss the photon dispersion relations in the photon gas,
which follow from the pole of the resummed photon propagator.
In Coulomb gauge the resummed propagator is given by
\bea
D_L^{-1}(p_0,p)&=&p^2-\Pi_L(p_0,p)=(1+\gamma)\> p^2,\nonumber \\
D_T^{-1}(p_0,p)&=&p_0^2-p^2-\Pi_T(p_0,p)=(1+\gamma)\> p_0^2-(1-\gamma)
\> p^2.
\label{eq13}
\eea
Whereas there is no dispersion relation for longitudinal photons,
i.e., there are no plasmons in the  photon gas, the dispersion relation of the
transverse (physical) photons is modified compared to the vacuum.
It is given by
\beq
\omega (p)=\sqrt{\frac{1-\gamma}{1+\gamma}}\> p\simeq (1-\gamma)\> p,
\label{eq14}
\eeq
i.e., the dispersion is located below the light cone $\omega < p$ and the 
plasma frequency $\omega_{pl}=\omega (p=0)$ vanishes.
The phase velocity $v_p=\omega /p\simeq 1-\gamma$ 
is identical to the group velocity $v_g=\partial \omega /\partial p$
and smaller than the speed of light in the vacuum. The result agrees
with \cite{Lat95}, where it has been derived in QED. Note also that
the phase as well as the group velocity are independent of the momentum, i.e.,
there is no dispersion.

Finally we turn to the dielectric tensor. In an isotropic medium there are
only two independent components of the dielectric tensor, for which we choose 
the longitudinal and the transverse dielectric functions,
related to the photon self energy via
\cite{Elz89}
\bea
\epsilon_L(p_0,p)&=&1-\frac{\Pi_L(p_0,p)}{p^2}=1+\gamma,\nonumber \\
\epsilon_T(p_0,p)&=&1-\frac{\Pi_T(p_0,p)}{p_0^2}
=1+\gamma \> \frac{p_0^2+p^2}{p_0^2}.
\label{eq15}
\eea
Since there is no direction preferred for ${\bf p}=0$ \cite{Lif81}, 
the longitudinal and 
transverse dielectric functions coincide in this limit,
$\epsilon_L(p_0,p=0)=\epsilon_T(p_0,p=0)=1+\gamma $. 

The electric permittivity and the magnetic permeability given by
\cite{Wel82}
\bea
\epsilon &=& \epsilon_L=1+\gamma,\nonumber \\
\frac{1}{\mu}&=&1+\frac{\Pi_T-p_0^2 \Pi_L/p^2}{p^2}=1-\gamma 
\label{eq16}
\eea
are independent of $p_0$ and $p$. The phase velocity following from
\cite{Wel82}, related to the index of refraction $n$,
\beq
v_p=\frac{1}{n}=\frac{1}{\sqrt{\mu \epsilon}}\simeq 1-\gamma
\label{eq17}
\eeq
agrees with the result found from the dispersion relation (\ref{eq14}).        

Owing to the cosmic microwave background the velocity of light in the 
Universe is reduced compared to the vacuum. Actually it increases  continuously
with time as the temperature drops. Today at a temperature of 2.7 K
it is given by (\ref{eq17}) with $\gamma = 4.7 \times 10^{-43}$. 
In the early Universe, when the radiation 
decoupled from matter at a temperature of about 3000 K we had 
$\gamma=6.5 \times 10^{-31}$.
Although this is probably not a measurable effect, the speed of light is 
not a constant in our Universe.

Summarizing, we have calculated the photon self energy in an
isotropic, non-equilibrium photon gas using the real time formalism.
For this purpose we considered the effective Lagrangian for photon-photon
interaction and calculated the photon self energy to lowest order
perturbation theory using an effective 4-photon vertex in momentum space.
As physical consequences of this self energy we showed the absence
of Debye and static magnetic screening in the photon gas. Also there are
no longitudinal collective modes (plasmons). However, the transverse collective
modes exhibit a modified dispersion with a vanishing plasma frequency and a
smaller slope compared to the vacuum modes. This results in a reduced,
dispersion free velocity of light, which increases during the evolution of 
the Universe as the temperature of the cosmic microwave background drops. 
We also determined the dielectric tensor, the electric permittivity and the
magnetic permeability of the photon gas. 
\vspace*{0.5cm}

\centerline{\bf ACKNOWLEDGMENTS}
\vspace*{0.3cm}
The author would like to thank U. Mosel and A. Rebhan
for stimulating and helpful discussions.

\vspace{-0.5cm}

\begin{figure}
\centerline{\psfig{figure=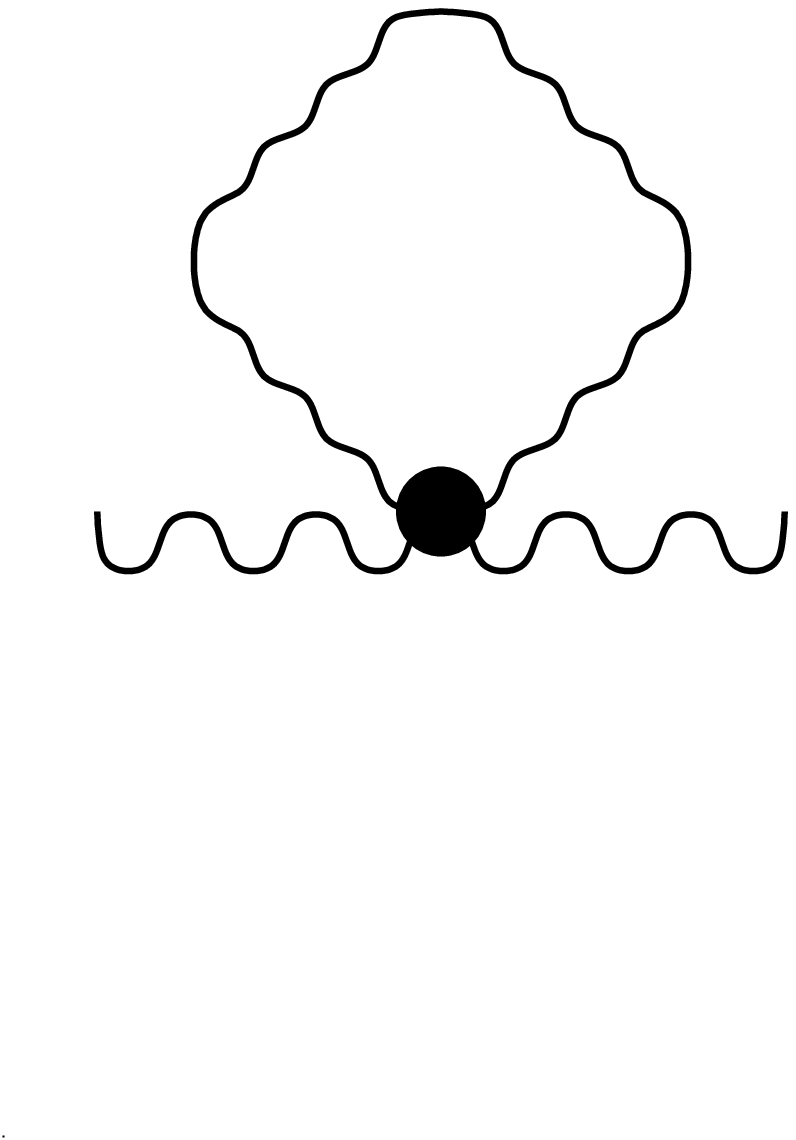,width=9cm}}
\vspace*{-5cm}
\caption{Lowest order photon self energy in the effective theory for
photon-photon interaction. The blob denotes the effective 4-photon vertex 
given in Ref.[3].}
\end{figure}


\begin{references}
\bibitem{Eul36} O. Halpern, Phys. Rev. 44 (1934) 855; H. Euler and B. Kockel,
Naturwiss. 23 (1935) 246; H. Euler, Ann. Phys. 26 (1936) 398;
W. Heisenberg and H. Euler, Z. Phys. 98 (1936) 714.
\bibitem{Sch51} J. Schwinger, Phys. Rev. 82 (1951) 664.
\bibitem{Hal93} J. Halter, Phys. Lett. B 316 (1993) 155.
\bibitem{Kap89} J.I. Kapusta, {\it Finite Temperature Field Theory} (Cambridge
University Press, New York 1989).
\bibitem{Wol83} R.M. Woloshyn, Phys. Rev. D 27 (1983) 1393.
\bibitem{Kon98} X. Kong and F. Ravndal, Nucl. Phys. B 526 (1998) 627.
\bibitem{Par98} C. Caso et al., Eur. Phys. J. C 3 (1998) 1.
\bibitem{Jac99} J.D. Jackson, {\it Classical Electrodynamics} (3rd ed., 
John Wiley, New York 1998).
\bibitem{Nie83} J.F. Nieves, P.B. Pal, and D.G. Unger, Phys. Rev. D 28 (1983)
908. 
\bibitem{Car99} M.E. Carrington, H. Defu, and M.H. Thoma, Eur. Phys. J C
7 (1999) 347.
\bibitem{Mro00} S. Mr\'owczy\'nski and M.H. Thoma, hep-ph/0001230 (to be 
published in Phys. Rev. D)
\bibitem{Dic98} D.A. Dicus, C. Kao, and W.W. Repko, Phys. Rev. D 57 (1998) 
2443.
\bibitem{Reb93} A.K. Rebhan, Phys. Rev. D 48 (1993) R3967. 
%\bibitem{Wan96} E. Wang and U. Heinz, Phys. Rev. D 53 (1996) 899.
\bibitem{Lat95} J.I. Latorre, P. Pascual and R. Tarrach, Nucl. Phys. B 437 
(1995) 60.
\bibitem{Elz89} H.T. Elze and U. Heinz, Phys. Rep. 183 (1989) 81.
\bibitem{Lif81} E.M. Lifshitz and L.P Pitaevskii, {\it Physical Kinetics}
(Pergamon Press, Oxford 1981).
\bibitem{Wel82} H.A. Weldon, Phys. Rev. D 26 (1982) 1394.
\end{references}
\end{document}